\documentclass[11pt]{article}
\usepackage{amsmath,amssymb,color,graphics,epsfig,cite}
\usepackage{graphicx,subfigure}

\usepackage{amsfonts}

\newcommand{\hoch}[1]{$\, ^{#1}$}

\newcommand{\be}{\begin{equation}}
	\newcommand{\ee}{\end{equation}}
\newcommand{\bea}{\setlength\arraycolsep{2pt} \begin{eqnarray}}
	\newcommand{\eea}{\end{eqnarray}}
\newcommand{\nn}{\nonumber}

\def\ft#1#2{{\textstyle{\frac{\scriptstyle #1}{\scriptstyle #2} } }}
\def\fft#1#2{{\frac{#1}{#2}}}

\def\0{{\sst{(0)}}}
\def\1{{\sst{(1)}}}
\def\2{{\sst{(2)}}}
\def\3{{\sst{(3)}}}
\def\4{{\sst{(4)}}}
\def\5{{\sst{(5)}}}
\def\6{{\sst{(6)}}}
\def\7{{\sst{(7)}}}
\def\8{{\sst{(8)}}}
\def\sst#1{{\scriptscriptstyle #1}}

\def\ep{\epsilon}

\def\pd{\partial}

\def\dd{\mathrm{d}}
\def\ext{\text{ext}}
\def\tr{\tilde{r}}

\begin{document}
	
	\begin{center}
		{\Large {\bf Black Hole Entropy Bounded by the Specific Heat}}
		
		\vspace{20pt}
		
		Kai-Peng Lu\hoch{1} and H. L\"{u}\hoch{1,2}
		
		\vspace{10pt}
		
		{\it \hoch{1}Center for Joint Quantum Studies, Department of Physics,\\
			School of Science, Tianjin University, Tianjin 300350, China }
		
		\medskip
		
		{\it \hoch{2}The International Joint Institute of Tianjin University, Fuzhou,\\ Tianjin University, Tianjin 300350, China}
		
		\vspace{40pt}
		
		\underline{ABSTRACT}

	\end{center}

We propose new bounds on black hole entropy in terms of the specific heat at fixed charges. For special spherically-symmetric and static black holes ($g_{tt} g_{rr}=-1$), we prove the bounds with suitable sufficient energy conditions. For more general static and rotating black holes, we test the bounds with a variety of examples. This work extends the previously-known algebraic inequalities of the black hole thermodynamic variables such as the Penrose inequality, to include their derivatives.

\vfill {\footnotesize kaipenglu@tju.edu.cn \ \ \ mrhonglu@gmail.com}
	
	\thispagestyle{empty}
	\pagebreak
	
	

\section{Introduction}
\label{sec:introduction}

The confluence of general relativity and quantum mechanics has revealed that black holes behave as thermodynamic objects \cite{Hawking:1971vc,Hawking:1975vcx}. The formulation of the four laws of black hole mechanics \cite{Bardeen:1973gs} provides a framework in which black holes are endowed with a temperature and an entropy. The Bekenstein-Hawking entropy, which equates the entropy to a quarter of the event horizon's area, forges a profound connection between geometry, information, and thermodynamics \cite{Bekenstein:1973ur}.

As with any thermodynamic system, a central question concerns the stability of a black hole. The global stability is related to the free energy compared to the vacuum thermal bath. The local stability is governed by the specific heat capacity at constant charges, $C_{Q_i} = T(\pd S / \pd T)_{Q_i}$, with $Q_i$ representing all conserved quantities including electromagnetic charges and angular momenta. The celebrated case of the Schwarzschild black hole, which possesses a negative specific heat, demonstrates that it is thermodynamically unstable and evaporates via Hawking radiation \cite{Hawking:1975vcx} with increasing temperature. In contrast, certain black holes, such as large Reissner-Nordstr\"om (RN) black holes or those in anti-de Sitter (AdS) space, can have a positive specific heat, allowing them to achieve a state of stable thermal equilibrium with their surroundings \cite{Hawking:1982dh}. More interestingly, this stability gives rise to a rich landscape of thermodynamic phase transitions \cite{Hawking:1982dh,Chamblin:1999hg,
Chamblin:1999tk,Kubiznak:2016qmn, Gunasekaran:2012dq,Altamirano:2014tva}. The points where the specific heat capacity diverges signal a second-order phase transition, typically between a small, unstable black hole and a large, stable one. This divergence marks the boundary between phases, with the change in the sign of the heat capacity across this critical point determining the stability of the different black hole states \cite{Wei:2019yvs}. Therefore, the sign of the specific heat is a critical indicator of a black hole's physical behavior and its ultimate fate.

Beyond thermodynamic stability, the physical viability of a black hole is also constrained by the weak cosmic censorship conjecture (WCCC) \cite{Penrose:1969pc}. Following the perturbative approach in gedanken experiments first proposed by Sorce and Wald \cite{Sorce:2017dst,Wald:2018xxi} for the Kerr-Newman black hole, our recent model-independent analysis \cite{Wu:2024ucf,Lu:2025ntu} establishes that for extremal and near-extremal black holes, the validity of the WCCC hinges upon the sign of a single quantity, defined at zero temperature as
\be
W=\Big(\fft{\pd S}{\pd T}\Big)_{Q_i;T=0}\,.
\ee
A positive value, $W>0$, ensures the WCCC is upheld at all orders \cite{Lu:2025ntu}. This condition links the WCCC to thermodynamic stability, since $W = \lim_{T\to 0} (C_{Q_i}/T)$. This naturally motivates a search for a universal bound related to stability for all black holes, not just the extremal or the near-extremal ones. To do so, we must consider the finite-temperature generalization of $W$, which is simply the quantity $(\pd S / \pd T)_{Q_i} = C_{Q_i}/T$. However, this quantity is ill-suited for establishing a universal bound, as it diverges wherever the specific heat capacity $C_{Q_i}$ does, namely at second-order phase transitions. To construct a well-behaved quantity that is finite across phase transitions, we instead consider its inverse:
\be
	U \equiv \Big(\fft{\pd T}{\pd S}\Big)_{Q_i} = \fft{T}{C_{Q_i}}\,.
\ee
This inverse quantity preserves the crucial sign information that indicates thermodynamic stability, but it remains finite and well-defined, approaching zero at critical points. It is therefore a robust candidate for investigating universal bounds on black hole thermodynamics.

In this paper, we propose a sequence of new thermodynamic inequalities governing four-dimensional black holes:
\be
-\ft12 T/S \le U \le \alpha S^{-3/2}\,,
\label{thermo_inequality4D}
\ee
where $\alpha$ is a dimensionless parameter of order unity whose value depends on the class of the black hole solutions. For the solutions examined in this work, we find
\bea
\hbox{Spherically-symmetric and static}:\qquad &&\qquad \alpha = \fft{1}{4\sqrt\pi}\,,\cr
\hbox{Rotating}:\qquad &&\qquad \alpha = \fft1{2\sqrt{2\pi}}\,.\label{d4alpha}
\eea
Note that we always have an option of choosing the larger $\alpha$ if we do not care to distinguish the fine detail. The sequence of inequalities in \eqref{thermo_inequality4D} implies a deeper relationship between a black hole's entropy and its specific heat. For thermodynamically unstable black holes $\big(C_{Q_i}<0\big)$, it implies an upper bound on the magnitude of the entropy in terms of the specific heat capacity. For stable black holes $\big(C_{Q_i}>0\big)$, it provides a different upper bound on the entropy. In both cases, the entropy is bounded above, with
\be
\begin{cases}
S \le -\fft12 C_{Q_i}\,, \qquad\qquad\quad  C_{Q_i}<0\,,  \\
S \le \Big(\alpha\, T^{-1} C_{Q_i}\Big)^{\fft23}\,, \qquad  C_{Q_i}>0\,.
\label{bound}
\end{cases}
\ee
Note that the entropy bounds involve only the horizon quantities, independent of the mass, which is a variable determined asymptotically. For some special classes of black holes, we are able to prove the bounds with suitable sufficient energy conditions. For more general classes, we verify this conjecture against a wide range of black hole solutions in both asymptotically flat and (A)dS spacetimes. In all cases considered, the inequality \eqref{thermo_inequality4D} holds. This provides strong evidence that a black hole's entropy and specific heat capacity are not independent, but are constrained by fundamental bounds. The structure of this inequality, and its generalization to higher dimensions, suggests a deep and previously unexplored connection between thermodynamic stability, information, and the geometric properties of black holes.

This paper is organized as follows. In Sec.~\ref{sec:SS}, we consider $D=4$ spherically-symmetric and static black holes. We first prove the inequality \eqref{thermo_inequality4D} in two special cases. One is the extremal limit. The other is the class of special spherically-symmetric black holes with $g_{tt}g_{rr}=-1$. For the more general static black holes, we confirm the inequalities by some specific nontrivial examples. In Sec.~\ref{sec:rotating}, we confirm the inequalities by examining a variety of rotating black holes. In Sec.~\ref{sec:higher}, we aim to generalize \eqref{thermo_inequality4D} to higher dimensions, including both spherically symmetric black holes and rotating black holes. We conclude the paper in Sec.~\ref{sec:conclusion}. In order not to interrupt the flow of our main text, we present some technical details in the appendix.

\section{Spherically-symmetric and static black holes}
\label{sec:SS}
	
\subsection{Setup}

We begin with a general class of spherically-symmetric and static black holes of mass $M$ and conserved charges $Q_i$ in $D=4$ dimensions. The general metric takes the form
\be
\dd s^2 = h \dd t^2 + f^{-1} \dd r^2 +r^2 \dd \Omega_2^2\,, \qquad h= -e^{2 \chi} f\,,
\label{metric1}
\ee
where $f$ and $\chi$ are functions of $(r,M, Q_i)$. The event horizon $r_+$ is the largest root of $f(r_+, M(r_+), Q_i) = 0$. Here we treat the charges as independent variables, while the mass can be a function of the horizon radius $r_+$. The temperature and entropy are
\be
T = \fft{e^\chi f'(r, M(r_+), Q_i)}{4 \pi} \Big|_{r=r_+}, \qquad S = \pi r_+^2\,,
\label{TS-shperical}
\ee	
where a prime denotes a derivative with respect to $r$. Note that in this paper, $X'(r_+)$ is understood as $X'(r)|_{r\rightarrow r_+}$. Using the chain rule, it is straightforward to find that
\cite{Wu:2024ucf}
\be
U= \fft{e^{\chi(r_+)}}{16\pi^2 r_+} \Big[ 2f''+2f'\chi'+ r f' \pd_M \big( e^\chi f' \big) \Big]_{r=r_+}.
\ee
Here we used the thermodynamic identity that $ \ft{\dd M}{\dd r_+} = 2 \pi r_+ T $. The inequality \eqref{thermo_inequality4D} can thus be transformed into the purely geometric inequalities on the horizon
\be
-\fft1{2 r_+} e^{\chi(r_+)} f'(r_+) \le \fft{e^{\chi(r_+)}}{4} \big[ 2f''+2f' \chi' + r f' \pd_M \left( e^\chi f' \right) \big]_{r=r_+} \le \fft1{r_+^2}\,.\label{inequ-gen-spher}
\ee

\subsection{Proofs in special cases}

We now employ energy conditions to examine the inequalities in \eqref{inequ-gen-spher}. In this paper, we consider only Einstein gravity with minimally coupled matter. For spherically-symmetric and static black holes, the matter energy-momentum tensor is diagonal and takes the form $ T^{\mu}_{\nu} = \text{diag}\{-\rho(r), p_r(r), p_T(r), p_T(r)\} $. The Einstein field equations reduce to
\bea
\rho &=& \frac{1}{8\pi r^2} \big(1-f-rf' \big)\,, \nn \\
p_r &=& \frac{1}{8\pi r^2} \Big(rf'+f \big( 1 + 2 r \chi' \big) -1 \Big) \,, \nn \\
p_T &=& \frac{1}{16\pi r} \Big(f' \big(2 + 3r \chi'\big) + r f'' + 2 f \big(\chi' + r \chi'^2 + r \chi''\big) \Big)\,.
\eea
Various energy conditions may provide strong constraints on the metric functions. For example,
the NEC $\rho + p_r\ge 0$ implies that $\chi' \ge 0$. However, we find that the energy conditions provide no general information about the term $\pd_M \big(e^\chi f'\big)$. We therefore prove \eqref{inequ-gen-spher} in two special cases.

\subsubsection{Extremal black holes}

In the extremal limit, $f(r_+)=f'(r_+)=0$. The SEC $\rho + p_r + 2 p_T\ge 0$ on the horizon implies that $f''(r_+)\ge 0$, which ensures the lower bound of \eqref{inequ-gen-spher} holds trivially. (It implies that the WCCC is valid in the gendanken experiments \cite{Wu:2024ucf,Lu:2025ntu}.)  Moreover, since $\chi' \ge0$ and $e^\chi \rightarrow 1$ at the asymptotic spatial infinity, we have
\be
\big(e^\chi f'' \big)_{r=r_+} \le f''(r_+)\,.
\ee
Then DEC ($\rho-p_T\ge 0$) on the horizon implies $f''(r_+) \le \ft{2}{r_+^2}$. Therefore, when DEC holds on the horizon, the upper bound of \eqref{inequ-gen-spher} also holds for general extremal black holes.

\subsubsection{Special static black holes}

The special static black holes with spherical symmetry are those with $g_{tt}g_{rr}=-1$, corresponding to taking $\chi=0$. The inequalities in \eqref{inequ-gen-spher} reduce to
\be
-\fft1{r_+} f'(r_+) \le \Big[ f'' + \fft{r}2 f' \big(\pd_M f'\big)\Big]_{r=r_+} \le \fft2{r_+^2} \,.
\label{chi0}
\ee
As shown in appendix~\ref{app:ansatz f}, the function $f$ for special static black holes has, in general, the form
\be
f(r,M,Q) = 1 - \fft{2M}{r} + g(r,Q_i)\,,\label{ftype}
\ee
for which $\big(\pd_M f'\big)_{r=r_+}=2/{r_+^2}$. Therefore, the sequence of inequalities in \eqref{chi0} can be written as
\be
-\fft{1}{r_+} f'(r_+) \le \Big( f''+\fft1{r} f'\Big)_{r=r_+} \le \fft2{r_+^2} \,.
\label{explicitineq}
\ee
It can be easily verified that the middle term above is given by
\bea
\Big( f''+\fft1{r} f'\Big)_{r=r_+} &=& -\fft{1}{r_+} f'(r_+) + 8\pi (\rho + p_r + 2p_T)\Big|_{r=r_+}\,,\label{cond1}\\
\Big( f''+\fft1{r} f'\Big)_{r=r_+} &=& \fft{2}{r_+^2}  - \fft{3f'(r_+)}{r_+}- 16\pi (\rho-p_T)\Big|_{r=r_+}\,.\label{cond2}
\eea
Thus, SEC ensures the lower bound. Furthermore, it follows from $f'(r_+)\geq 0$ that if a special static black hole has $g_{tt}g_{rr} = -1$ and satisfies the DEC, then it must satisfy the upper bound of \eqref{explicitineq}. Note that the Schwarzschild black hole saturates the lower bound of \eqref{thermo_inequality4D}, and the extremal RN black hole saturates the upper bound.

It should be emphasized that for the inequalities to hold, the required energy conditions need to be imposed on the horizon only. It follows from \eqref{cond1} the lower bound is saturated when the SEC is saturated and the violation of the SEC on the horizon necessarily violates the lower bound. The saturation of the upper bound, on the other hand, requires both the extremality condition $f'(r_+)=0$ and the saturation of the DEC $\rho = p_T$. Thus, for non-extremal black holes, the upper bound can hold even when the DEC is relaxed, as long as it does not overpower the last term in \eqref{cond2}.

The sequence of inequalities in \eqref{bound} neatly partitions the parameter space of black holes into stable and unstable phases. The boundaries defined by saturating \eqref{bound} correspond to extremal objects in their respective classes: the Schwarzschild solution in the unstable regime, and a specific power-law abiding black hole in the stable one. This suggests the inequalities \eqref{bound} are fundamental constraints on the endpoints of gravitational collapse and black hole evolution within each stability class. However, a general proof using only energy conditions on the abstract metric \eqref{metric1} appears intractable due to the unknown behavior of the term $\pd_M \left( e^\chi f' \right)$. Nevertheless, we shall next study a variety of examples to show that \eqref{thermo_inequality4D} still holds for more complex solutions, including rotating black holes and those in spacetime with a non-zero cosmological constant.

Before ending of this subsection, we note that the NEC $(\rho+p_T\ge0)$ alone also provides a lower bound, namely
\be
-16\pi \rho(r_+) - \fft{1}{r_+}f'(r_+) = - \fft{2}{r_+^2} + \fft{1}{r_+}f'(r_+) \le  \Big( f''+\fft1{r} f'\Big)_{r=r_+}\,.
\label{explitineq}
\ee
This is a less restrictive lower bound when the WEC is satisfied.

\subsection{Explicit examples}

In the previous subsections, we established for spherically-symmetric and static black holes with $\chi=0$, the inequalities of \eqref{thermo_inequality4D} reduce to those in \eqref{explicitineq}. We showed that the SEC and DEC were sufficient for the lower and upper bounds for the special static black holes. For the more general $\chi\ne 0$ black holes, a proof is absent and we shall test the inequalities with some explicit examples. We shall also consider the special static examples to clarify a few points.

In this subsection, we test our proposed sequence of inequalities \eqref{thermo_inequality4D} against some nontrivial black hole solutions. The lower and upper bounds of \eqref{thermo_inequality4D} can be transformed into following two inequalities respectively
\bea
Y_1 &\equiv& U + \ft12 T/S \geq 0\,,  \label{lowbound}\\
Y_2 &\equiv& U - \alpha S^{-3/2} \le 0\,, \label{upbound}
\eea
where the numerical constant $\alpha$ was specified by \eqref{d4alpha}.

\subsubsection{RN-(A)dS black holes}

The RN black hole satisfies all the energy conditions and hence it provides the simplest nontrivial example to illustrate both the upper and lower bounds in \eqref{thermo_inequality4D}. We add the cosmological constant $\Lambda$ to illustrate how SEC and DEC affect the bounds. A positive cosmological constant satisfies the DEC, whilst a negative cosmological constant satisfies the SEC. Thus we expect that the lower bound will be upheld by the RN-AdS black hole, whilst the upper bound will be upheld by the RN-dS black hole. To see this explicitly, we note that
\be
f=-\ft13 \Lambda r^2 + 1 - \fft{2m}{r} + \fft{Q^2}{r^2}\,.
\ee
Thus, we have
\be
T=\fft{f'(r_+)}{4\pi}\,,\qquad S=\pi r_+^2\,,\qquad U = \fft{3Q^2 - r_+^2 - r_+^4 \Lambda}{8\pi^2 r_+^5}\,,
\ee
and
\be
Y_1 = \fft{Q^2 - r_+^4\Lambda}{4\pi^2 r_+^5}\,,\qquad Y_2 = - \fft{3T}{2\pi r_+^2} - \fft{\Lambda}{2\pi^2 r_+}\,.
\ee
It is clear that $\Lambda\le 0$ ensures that $Y_1> 0$, whilst $\Lambda\ge 0$ ensures that $Y_2< 0$. For vanishing $\Lambda$, the lower bound is saturated by the Schwarzschild black hole $(Q=0)$ and the upper bound is saturated by the extremal RN black hole.

\subsubsection{Bardeen black holes}

Bardeen black hole is a regular black hole satisfying the WEC throughout the spacetime.  The metric is of the special static type, given by \eqref{metric1}, with
\be
h=f=1 - \fft{2M r^2}{(r^2 + q^2)^{\fft32}}\,.\label{Bardeen1}
\ee
We include the Bardeen black hole as a specific example for discussion since it involves two subtleties. One is that any regular black hole necessarily violates the SEC, and it further violates the DEC. Furthermore the mass $M$ does not fit the general pattern of the type \eqref{metric1}, assumed in the general proof. The temperature, entropy and the function $U$ of the Bardeen black hole are
\be
T=\fft{r_+^2 -2 q^2}{4 \pi r_+ \big(q^2+r_+^2\big)}\,, \qquad S=\pi r_+^2\,, \qquad U=\fft{2 q^4+7 q^2 r_+^2-r_+^4}{8 \pi ^2 r_+^3 \big(q^2+r_+^2\big)^2}\,.
\label{Bard-TSU1}
\ee
Typically the black hole have two horizons, which coalesce in the extremal limit.  The mass and the horizon radius of the extremal solution are
\be
M_{\rm ext}= \ft34\sqrt3\,q\,,\qquad r_{\rm ext}=\sqrt2\,q\,.
\ee
For given $q$, the general black hole has $M\ge M_{\rm ext}$ with the outer horizon $r_+\ge r_{\rm ext}$. The quantities $Y_1$ and $Y_2$ take the following forms
\be
Y_1 = \fft{3 q^2}{4 \pi ^2 r_+ \big(q^2+ r_+^2 \big)^2}\,, \qquad
Y_2 =- \fft{3 \big(r_+^2-q^2\big)}{8 \pi ^2 r_+ \big(q^2+r_+^2\big)^2}\,.
\ee
It is easy to see that both \eqref{lowbound} and \eqref{upbound} hold. However, it is worth pointing out that \eqref{Bardeen1} is not the most general solution for the given matter energy-momentum tensor, since we can always add a term constant$/r$ associated with the condensation of massless graviton in the above and the metric will satisfy the exact same equations of motion \cite{Li:2023yyw,Huang:2025uhv}. It is interesting to examine whether this general solution satisfies our thermodynamic inequalities of \eqref{thermo_inequality4D}. The metric of this more general solution is given by
\be
h=f=1 -\fft{2(m-M)}{r} -\fft{2 M r^2}{\big(r^2+q^2\big)^{3/2}}\,,
\label{Bardeen2}
\ee
where $m$ is the free mass parameter, and $(M,q)$ are fixed parameters associated with the matter energy-momentum tensor. For the specific choice of mass $m=M$, this metric reduces to \eqref{Bardeen1} and the singularity at $r=0$ disappears.

The temperature, entropy and $U$ of the general Bardeen metric are
\bea
T &=& \frac{\left(q^2+r_+^2\right)^{5/2}-6 M q^2 r_+^2}{4 \pi  r_+ \left(q^2+r_+^2\right)^{5/2}}\,, \qquad S=\pi r_+^2\,,\cr
U\Big|_{M,q}&=&\frac{-6 M q^4 r_+^2+24 M q^2 r_+^4-\left(q^2+r_+^2\right)^{7/2}}{8 \pi ^2 r_+^3 \left(q^2+r_+^2\right)^{7/2}}\,.
\label{Bard-TSU2}
\eea
By setting $m=M$, we have $M=\left(q^2+r_+^2\right)^{3/2}/(2 r_+^2)$, the same as the one derived from \eqref{Bardeen1}. Substituting this relation into \eqref{Bard-TSU2}, we can find that both temperature and entropy are the same as \eqref{Bard-TSU1}, but $U$ is different. Therefore, it is worth checking whether the general Bardeen metric satisfies the inequality \eqref{thermo_inequality4D} after taking $m=M$. We find now that $Y_1$ and $Y_2$ take the following forms
\be
Y_1 = \frac{3 \left(3 q^2 r_+^2-2 q^4\right)}{8 \pi ^2 r_+^3 \left(q^2+r_+^2\right)^2}\,, \qquad
Y_2 = -\frac{3 \big[ ( r_+^2 - q^2 )^2 +q^4\big]}{8 \pi ^2 r_+^3 \left(q^2+r_+^2\right)^2}\,.
\ee
Therefore, $Y_1$ and $Y_2$ for \eqref{Bardeen2} are also different from those for the metric \eqref{Bardeen1}, but both \eqref{lowbound} and \eqref{upbound} still hold. We have examined a variety of electrically-charged regular black holes \cite{Li:2024rbw} and find the conclusion is the same.

We now comment on the issues of the energy conditions which were essential in the general proof of the inequalities for the special static black holes. Although the Bardeen black hole necessarily violates the SEC, the SEC on the horizon actually hold is satisfied:
\be
\rho + p_r + 2 p_T = 2p_T = \fft{3q^2(3r_+^2 - 2q^2)}{8\pi r_+^2 (q^2 + r_+^2)^2} >0\,.
\ee
Thus the lower bound $Y_1> 0$ is not surprising. On the other hand, the DEC on the horizon can be violated with sufficiently large black hole, namely
\be
\rho - p_T = - \fft{3q^2(r_+^2 - 4q^2)}{16\pi r_+^2 (r_+^2 + q^2)^2}\le 0 \,, \qquad\hbox{for}\qquad r_+\ge 2q\,.
\ee
This illustrates that the required energy conditions in the general proof are sufficient conditions and hence are not always necessary. To be specific, in the extremal limit, the DEC is necessary and indeed DEC is satisfied on the horizon for the extremal Bardeen black hole. For $r_+\ge 2 q$, the black hole becomes sufficiently non-extremal and DEC becomes unnecessary.

\subsubsection{Einstein-Maxwell dilaton theories}

For the more general $h\ne f$ black holes, we do not have a proof of the inequalities; therefore, it is necessary to test examples in literature. Here we present the explicit demonstration for the charged black holes in Einstein-Maxwell-dilaton (EMD) theory. The Lagrangian is
\be
{\cal L}= \sqrt{-g} (R - \ft12 (\pd \phi)^2 - \ft14 e^{a\phi} F^2)\,,\qquad F=\dd A\,,\qquad a^2=\fft{4}{N} - 1\,.
\ee
The parameter $N$ can take values within $(0,4]$. When $N=1,2,3,4$, the solutions can be embedded in supergravities and lifted to become black branes or intersecting branes in string or M-theory \cite{Cvetic:1996gq,Duff:1996hp}. The metric of the black hole solution is
\bea
\dd s^2 &=& -H^{-\fft12 N} \tilde f \dd t^2 + H^{\fft12 N} \Big(\fft{\dd r^2}{\tilde f} + r^2 \dd\Omega_2^2\Big)\,,\nn\\
\tilde f &=& 1- \fft{\mu}{r}\,,\qquad H=1 + \fft{q}{r}\,.
\label{singly-charged}
\eea
The black hole mass, charge and the horizon radius are given by
\be
M=\ft12 \mu + \ft14 N q\,,\qquad Q=\ft14 \sqrt{N} \sqrt{q(\mu+q)}\,, \qquad  R_+ = r_+ \Big(1 + \fft{q}{r_+}\Big)^{\fft14 N}\,,
\ee
where $r_+=\mu$. The temperature and entropy of the black hole are
\be
T=\ft1{4\pi} H(r_+)^{-\fft12 N} \tilde f'\big(r_+\big)\,, \qquad S=\pi R_+^2\,.
\label{TS-singly}
\ee

Here we simply consider the cases that $N=1,\, 2,\, 3,\, 4$. The corresponding $Y_1$ and $Y_2$ quantities are listed in Table.~\ref{table.1}. For $N=1, 2$ and 4, it is easy to see that both \eqref{lowbound} and \eqref{upbound} hold. For $N=3$, it is also clear that \eqref{lowbound} holds. For the upper bound \eqref{upbound}, we can prove that it holds by setting $q=x \mu$, where $x \in (0,\infty)$. The numerator of $Y_2$ is transformed to
\be
-\mu ^2 (x+1)^{1/4} \left((x+2) (x+1)^{3/4}-x+1\right),
\ee
which is less than 0. The general $N \in (0,4)$ case is presented in appendix~\ref{app:gen-N}, where we use a semi-analytical analysis to show that both \eqref{lowbound} and \eqref{upbound} hold.

\begin{table}[tbh]
	\centering
	\tabcolsep=6mm
	\renewcommand\arraystretch{2}
		\begin{tabular}{c|c|c}
			\hline
			\hline
			$N$ & $Y_1$ & $Y_2$ \\
			\hline			
			1 & \Large $\frac{q}{8 \pi ^2 \mu ^2 (\mu +q) (2 \mu +3 q)}$ & \Large $-\frac{2 \mu +\mu^{1/4} (\mu +q)^{3/4} +3 q}{4 \pi ^2 \mu ^{9/4} (\mu +q)^{3/4} (2 \mu +3 q)}$ \\
			\hline
			2 & \Large $\frac{q}{8 \pi ^2 \mu  (\mu +q)^3}$ & \Large $-\frac{2 \mu ^2+2 q^2+\mu ^{3/2} (\mu +q)^{1/2} +4 \mu q}{8 \pi ^2 \mu ^{3/2} (\mu +q)^{7/2}}$ \\
			\hline
			3 & \Large $\frac{3 q}{8 \pi ^2 (\mu +q)^3 (2 \mu +q)}$ & \Large $-\frac{2 \mu^2+q^2+3 \mu q + \mu ^{3/4} (\mu-q) (\mu +q)^{1/4}}{4 \pi ^2 \mu ^{3/4} (\mu +q)^{13/4} (2 \mu +q)}$ \\
			\hline
			4 & \Large $\frac{q}{4 \pi ^2 (\mu +q)^4}$ & \Large $-\frac{3 \mu }{8 \pi ^2 (\mu +q)^4}$ \\
			\hline
			\hline
		\end{tabular}
\caption{The quantities $Y_1$ and $Y_2$ for $N=1,\,2,\,3,\,4$ of the EMD black holes, for which both bounds \eqref{lowbound} and \eqref{upbound} are satisfied. \label{table.1}}
\end{table}

\section{Rotating black holes}
\label{sec:rotating}

We now consider the sequence of inequalities in \eqref{thermo_inequality4D} for rotating black holes in four dimensions. The simplest such example is the Kerr black hole. We find that the lower bound remains; however, the upper bound with $\alpha = 1/(4\sqrt\pi)$ is violated by the extremal Kerr black hole. Instead of declaring that there is no upper bound, we find that a bound still exists if we replace the $\alpha$ by a larger $\alpha=1/(2\sqrt{2\pi})$, in which case, the extremal Kerr black hole saturates the bound. This new $\alpha$ is $\sqrt2$ factor of the one established for the spherically-symmetric black holes.

For rotating black holes, we are unable to provide the similar energy-condition analysis for the spherically-symmetric case. This is because a general ansatz for rotating black holes are too complicated. At first sight, the quantities related to the inequalities are all horizon properties and hence only the horizon geometry is necessary. However, writing the most general horizon geometry alone will not enable us to distinguish the charge parameters, which should be fixed to evaluate the specific heat capacity or the quantity $U$.  In this section, we shall consider a variety of known exact solutions of rotating black holes and test the bounds \eqref{lowbound} and \eqref{upbound} with the new $\alpha$.

\subsection{Kerr black hole}

We first examine the Kerr black hole, whose metric is well established in literature and we shall not repeat here. The solution contains two integration constants, the mass $M$ and angular momentum $J$. The temperature and entropy are
\be
T=\fft{r_+ - M}{2\pi \big(r_+^2 +a^2 \big)}\,, \qquad S=\pi \big( r_+^2 +a^2 \big)\,,
\ee
where $a=J/M$ is the unit angular momentum and $r_+ =M+\sqrt{M^2-a^2}$ is the outer horizon, with $r_+ \geq r_\ext=a$. Both $M$ and $J$ can be written as functions of $r_+$ and $a$, and it is straightforward to find that
\be
Y_1= \frac{a^4+3 a^2 r_+^2}{4 \pi ^2 r_+ \left(a^2+r_+^2\right)^3}\,, \quad
Y_2= -\frac{-3 a^4+2 \sqrt{2} r_+ \left(a^2+r_+^2\right)^{3/2}-6 a^2 r_+^2+r_+^4}{8 \pi ^2 r_+ \left(a^2+r_+^2\right)^3}\,.
\ee
It is easy to see that \eqref{lowbound} holds automatically and the sign of $Y_2$ depends only on its numerator. We can denote $x=a/r_+$, where $x \in [0,1]$. The numerator of $Y_2$ is transformed as
\be
Z=3 x^4-2 \sqrt{2} \left(x^2+1\right)^{3/2}+6 x^2-1\,.
\ee
The derivative of $Z$ with respect to $x$ is $6 x \sqrt{2 (x^2+1)} \big(\sqrt{2 (x^2+1)}-1\big)$, which is greater than 0 at $x \in (0,1]$, indicating that $Z(x)$ is a monotonically-increasing function in the concerned $x$ region. Therefore, $Z_\text{max}=Z|_{x=1}=0$, which leads to $Y_2 \le 0$ at $x \in [0,1]$. Thus, \eqref{upbound} also holds for Kerr black holes, and extremal ones saturate this bound. In fact, as we have mentioned earlier, the parameter $\alpha$ was established precisely by the extremal Kerr black hole. In the next subsections,  we shall confirm the bounds with a variety of rotating black holes involving non-vanishing matter energy-momentum tensor.

\subsection{Kerr-Newman black hole}

We now consider the Kerr-Newman black hole of Einstein-Maxwell gravity as a nontrivial example. Since the electric $Q$ and magnetic charges $P$ enter the metric only via $Q^2 + P^2$, we shall set $P=0$ without loss of generality. The temperature and entropy of the Kerr-Newman black hole are
\be
T=\fft{r_+ - M}{2\pi \big(r_+^2 +a^2 \big)}\,, \qquad S=\pi \big( r_+^2 +a^2 \big)\,,
\ee
where $r_+=M+\sqrt{M^2-a^2-Q^2}$ is the outer horizon. Similar to the Kerr black hole, both $M$ and $J$ can be written as functions of $r_+,\, a$ and $Q$. It is straightforward to find that
\bea
Y_1 &=& \frac{a^4+a^2 \left(2 Q^2+3 r_+^2\right)+Q^2 \left(Q^2+r_+^2\right)}{4 \pi ^2 r_+ \left(a^2+r_+^2\right)^2 \left(a^2+Q^2+r_+^2\right)}\,, \\
Y_2 &=& \frac{Y_3}{8 \pi ^2 r_+ \left(a^2+r_+^2\right)^{5/2} \left(a^2+Q^2+r_+^2\right)}\,,
\eea
where
\be
Y_3= \sqrt{a^2+r_+^2} \Big[2 r_+^2 \big(3 a^2+Q^2\big)+3 \big(a^2+Q^2\big)^2-r_+^4\Big]-2 \sqrt{2} r_+ \big(a^2+r_+^2\big) \big(a^2+Q^2+r_+^2\big) \label{KNY3}\,.
\ee
Therefore, \eqref{lowbound} holds automatically and the sign of $Y_2$ is determined by the sign of $Y_3$.

In order to transform $Y_3$ into a neat form, we substitute $r_+=M+\sqrt{M^2-a^2-Q^2}$ into \eqref{KNY3} and set $M=1$ without loss of generality, so that the parameters $a$ and $Q$ become dimensionless. In order to avoid the naked singularity, we must have $a^2+Q^2 \le 1$. Therefore, we can introduce a quantity $x$ to simplify our result, where $x=1 + \sqrt{1 - Q^2 - a^2}$ and $1\le x \le 2$, such that we can replace $a$ by $a=\sqrt{-Q^2-(x-2) x}$. Now $Y_3$ is written as
\be
-4 x^2 \left(x^2 \sqrt{2 x-Q^2}+Q^2 \left(\sqrt{2 x-Q^2}-\sqrt{2}\right)-3 \sqrt{2 x-Q^2}+2 \sqrt{2} x\right)\,.
\label{KNY31}
\ee
In order to simplify this result further, we introduce two parameters $y$ and $z$, where $y=x-1$ and $z=1-Q^2$. Both $y$ and $z$ belong to $[0,1]$. Substituting these two parameters into the bracket term of \eqref{KNY31}, which becomes
\be
\left(y^2+2 y-z-1\right) \sqrt{2 y+z+1}+2 \sqrt{2} y+\sqrt{2} z+\sqrt{2}\,.
\label{KNY32}
\ee
Separating the positive terms and negative terms in \eqref{KNY32}, we can find that the square of the positive terms subtracting the square of the negative terms leads to the following result
\be
(2 y+z+1)\Big(y^4+4 y^3+\left(2 y^2+4 y\right) \big(\sqrt{2} \sqrt{2 y+z+1}+1\big)+2 y^2-z^2+1\Big)\,.
\ee
The positiveness of the above implies that \eqref{KNY32} must be positive and hence $Y_3$ must be negative. We thereby establish the upper bound \eqref{upbound} for the general Kerr-Newman black hole.

\subsubsection{Kerr-Sen black hole}
The Kerr-Sen black hole was generated from the Kerr black hole to carry an electric charge in the string theory \cite{Sen:1992ua}. The solution contains three parameters, the mass $M$, angular momentum $J$, and the electric charge $Q$. Temperature and entropy can be expressed as follows
\be
T= \frac{\Delta}{4 \pi  M \left(\Delta+2 M^2-Q^2\right)}\,, \qquad
S= \pi \left(\Delta+2 M^2-Q^2\right),
\label{KerrSenTS}
\ee
where $\Delta=\sqrt{\big(2M^2-Q^2 \big)^2-4J^2}$. To get a positive temperature, we must have $2M^2-Q^2 \geq 2J$, where we choose $J>0$ without loss of generality. $Y_1$ and $Y_2$ take the following forms
\bea
Y_1 &=& -\frac{2 M^2 \Delta-4 J^2-4 M^4+Q^4}{16 \pi ^2 M^3 \left(\Delta+2 M^2-Q^2\right)^2}\,, \label{KSY1}
\\
Y_2 &=& -\frac{4 M^2 \Delta+4 \sqrt{2} M^3 \sqrt{\Delta+2 M^2-Q^2}-4 J^2-4 M^4+Q^4}{16 \pi ^2 M^3 \left(\Delta+2 M^2-Q^2\right)^2}\,.
\label{KSY2}
\eea
The signs of $Y_1$ and $Y_2$ depend on the sign of their numerator. For simplicity, we can set $Q=\sqrt{2}q$ and measure $q$ and $J$ in unit of mass $M$, i.e.~we set $M=1$. The new constraint is given by $q^2+J \le 1$. First we consider the sign of the numerator of $Y_1$, which is written as
\be
Y_3=-4 \left(\sqrt{\left(1-q^2\right)^2-J^2}-J^2+q^4-1\right).
\ee
To simplify $Y_3$, we set $x_q=\left(1-q^2\right)^2-J^2$, where $0<x_q<1$. Therefore, $Y_3$ is rewritten as
\be
4 \left(2 \sqrt{J^2+x_q}-\left(x_q+\sqrt{x_q}\right)\right).
\ee
It is easy to see that $Y_3 \geq 0$, which leads to $Y_1 \geq 0$. Therefore, \eqref{lowbound} holds for Kerr-Sen black holes. Then we consider the upper bound \eqref{upbound}. Under the same choice of parameters $(Q=\sqrt{2}q,\, M=1)$, the numerator of $Y_2$ is written as
\be
Y_4=-4 \left(2 \sqrt{\left(1-q^2\right)^2-J^2}+2 \sqrt{\sqrt{\left(1-q^2\right)^2-J^2}-q^2+1}-J^2+q^4-1\right).
\ee
Similar to the simplification of $Y_1$, we set $x_J=\sqrt{\left(1-q^2\right)^2-J^2}$ and $q=\sqrt{1-z}$. Both $x_J$ and $z$ take values at $[0,1]$. Substituting $x_J$ and $z$ into $Y_4$, we find that
\be
Y_4=-4 \left(x_J^2+2 \sqrt{x_J+z}+2 x_J-2 z\right).
\ee
It is also easy to see that $Y_4 \le 0$ and hence \eqref{upbound} holds for general Kerr-Sen black holes.

\subsection{Kerr-(A)dS black hole}

As mentioned in Sec.~\ref{sec:SS}, we expect that lower bound should be valid for Kerr-AdS black hole, whilst the upper bound should be valid for the Kerr-dS black hole. In this subsection, we aim to establish this. The four-dimensional Kerr-AdS metric was first obtained by Carter \cite{Carter:1968ks}, which can be written as
\be
\dd s^2= -\fft{\Delta}{\rho^2} \Big[ \dd t - \fft{a}{\Xi}\sin^2 \theta \dd \phi \Big]^2 +\fft{\rho^2 \dd r^2}{\Delta} +\fft{\rho^2 \dd \theta^2}{\Delta_\theta} +\fft{\Delta_\theta \sin^2 \theta}{\rho^2} \Big[a \dd t - \fft{r^2+a^2}{\Xi} \dd \phi \Big]^2,
\ee
where
\bea
\Delta &=& \big( r^2+a^2 \big) \big( 1+r^2g^2 \big) -2mr\,, \quad \Delta_\theta= 1-a^2g^2 \cos^2 \theta\,, \nn \\
\rho^2 &=& r^2+a^2\cos^2 \theta\,, \qquad \Xi= 1-a^2g^2\,.
\label{KerrAdS}
\eea
The temperature and entropy are given by \cite{Gibbons:2004ai}
\be
T= \frac{r_+ \left(a^2 g^2-a^2 r_+^{-2} +3 g^2 r_+^2+1\right)}{4 \pi  \left(a^2+r_+^2\right)}\,, \qquad S= \fft{\pi \big(r_+^2 + a^2 \big)}{\Xi}\,,
\ee
where $r_+$ is the largest root of $\Delta=0$.  The mass and angular momentum are $M=m/\Xi^2$ and $J=Ma/\Xi^2$. We find
\bea
Y_1 &=& \fft{\big(1-a^2 g^2\big)}{8 \pi ^2 r_+ \big(a^2+r_+^2\big)^3} \Big[a^6 \left(g^4 r_+^2+g^2\right)+2 a^4 \left(3 g^4 r_+^4+7 g^2 r_+^2+1\right) \nn \\
&&+3 a^2 r_+^2 \left(3 g^4 r_+^4+9 g^2 r_+^2+2\right)+6 g^2 r_+^6\Big]\,.
\eea
The entropy must be positive, so $1-a^2g^2>0$. Therefore, $Y_1>0$ and the lower bound holds for Kerr-AdS black holes.

Now we move to Kerr-dS black holes and the metric of Kerr-dS black holes is the same as that of the Kerr-AdS black hole but replacing $g^2$ by $-g^2$ in \eqref{KerrAdS}. The temperature and entropy are similar, which are written as
\be
T= \frac{r_+ \left(1-a^2 g^2-a^2 r_+^{-2} -3 g^2 r_+^2\right)}{4 \pi  \left(a^2+r_+^2\right)}\,, \qquad S= \fft{\pi \big(r_+^2 + a^2 \big)}{1+a^2g^2}\,.
\label{TS-KerrdS}
\ee
Here, $r_+$ is still denoted as the outer event horizon. However, $r_+$ could not be too large, or we will get the cosmic horizon instead of the outer event horizon. $Y_2$ now takes the following form
\bea
Y_2 &=& \frac{1+a^2g^2}{8 \pi ^2 r_+ \big(a^2+r_+^2\big)^{7/2}} \bigg[ \sqrt{a^2+r_+^2} \Big(a^6 g^2 \big(g^2 r_+^2-1\big)+a^4 \big(6 g^4 r_+^4-13 g^2 r_+^2+3\big) \nn \\
&&+a^2 r_+^2 \big(9 g^4 r_+^4-23 g^2 r_+^2+6\big)-r_+^4 \big(3 g^2 r_+^2+1\big)\Big)-2 r_+ \sqrt{2 a^2 g^2+2} \big(a^2+r_+^2\big)^2 \bigg]\,.
\eea
The sign of $Y_2$ depends on the sign of the big bracket term, which we shall name it $Y_3$. The sign of $Y_3$ cannot be seen easily. We leave the details in appendix~\ref{app:KerrdS}, where we use semi-analytical method to prove that $Y_3<0$, which establishes the upper bound for the general Kerr-dS black holes.

\section{Generalizations to higher dimensions} \label{sec:higher}

In the previous sections, we have examined variety of black holes in four-dimensional spacetime. For some classes of black holes, we established the inequalities using strict proof whilst for others, we used specific examples to confirm the bounds. We now extend the discussions to higher-dimensional spacetime, and use the RN-Tangherlini black holes and Myers-Perry black holes to establish the bounds, if they exist.

\subsection{RN-Tangherlini black hole}

The RN-Tangherlini black hole was proposed by Tangherlini \cite{Tangherlini:1963bw} in 1963. The black hole metric is given by
\be
\dd s^2= -f \dd t^2 +f^{-1} \dd r^2 +r^2 \dd \Omega_{D-2}^2\,, \qquad f=1-\fft{2M}{r^{D-3}}+\fft{Q^2}{r^{2(D-3)}}\,,
\ee
where $\dd \Omega_{D-2}^2$ denotes the metric of a $(D-2)$-dimensional unit sphere $(D \geq 4)$. The temperature and entropy are given by
\be
T=\frac{(D-3) r_+^{-2 D-1} \left(r_+^{2 D}-Q^2 r_+^6\right)}{4 \pi }\,, \qquad
S=\ft14 r_+^{D-2} \Sigma_{D-2}\,,
\ee
where $\Sigma_D \equiv \fft{2\pi^{(D+1)/2}}{\Gamma \big((D+1)/2\big)}$ is the volume of the round unit $D$-sphere. In order to match the dimension of each quantity, the sequence of inequalities of \eqref{thermo_inequality4D} is now generalized to
\be
-\fft{1}{D-2} \fft{T}{S} \le U \le \frac{2^{\frac{D}{2-D}} (D-3)^2 \Sigma_{D-2}^{\frac{1}{D-2}} S^{\frac{1-D}{D-2}}}{\pi  (D-2)}\,.
\ee
For $D=4$, this sequence reduces to \eqref{thermo_inequality4D}. It is straightforward to find that
\bea
Y_1 &=& U+\fft{1}{D-2} \fft{T}{S}= \frac{2 (D-3)^2 Q^2 r_+^{7-3 D}}{\pi  (D-2) \Sigma_{D-2}}\,, \\
Y_2 &=& U-\frac{2^{\frac{D}{2-D}} (D-3)^2 \Sigma_{D-2}^{\frac{1}{D-2}} S^{\frac{1-D}{D-2}}}{\pi  (D-2)}
= -\frac{(D-3) (2 D-5) r_+^{1-3 D} \left(r_+^{2 D}-Q^2 r_+^6\right)}{\pi  (D-2) \Sigma_{D-2}}\,.
\eea
Thus, we see that $Y_1 \geq 0$, and $Y_2 \le 0$ since the temperature should be positive. It is also worth pointing that the Schwarzschild-Tangherlini black holes saturate the lower bound and the extremal RN-Tangherlini black holes saturate the upper bound. We expect that the bounds hold for spherically-symmetric and static black holes in higher dimensions, provided suitable energy conditions are satisfied.

\subsection{Myers-Perry black hole}

The Myers-Perry (MP) black holes \cite{Myers:1986un} are higher-dimensional generalizations of the Kerr metric in four dimensions, and they were generalized to higher-dimensional Kerr-AdS black holes in \cite{Hawking:1998kw,Gibbons:2004uw, Gibbons:2004js}. The black hole thermodynamic quantities of Kerr-AdS black holes in general dimensions were analysed in \cite{Gibbons:2004ai}. Here, we use the notation in \cite{Gibbons:2004ai}, but set zero the cosmological constant. In $D$ dimensions, there can be at most $N=[\ft{D-1}{2}]$ independent orthogonal rotations; therefore, the metrics contains $(N + 1)$ parameters, namely $(m,a_1,a_2,\cdots,a_N)$, parameterizing the mass and $N$ independent angular momenta:
\be
M= \fft{(D-2)\Sigma_{D-2}}{8\pi}m\,, \qquad J_i= \fft{\Sigma_{D-2}}{4\pi}m a_i\,, \qquad i=1,\,2,\,\cdots,\, N\,.
\ee
For sufficiently large mass $M$, there can exist an event horizon $r_+$, and the $N+1$ parameters and $r_+$ are related by the horizon condition
\be
m=\ft12 r_+^{\ep-2} \prod_i \big(r_+^2+a_i^2\big)\,, \qquad \ep=D-2N-1\,.
\ee
Here we introduce a notation $\ep$, which is 0 or 1 for $D=$ odd or even respectively. The corresponding temperature and entropy are given as
\be
T= \fft{1}{2\pi} \bigg( \sum_i \fft{r_+}{r_+^2+a_i^2} - \fft{1}{(\ep+1)r_+} \bigg)\,, \qquad S= \fft{\Sigma_{D-2}}{4 r_+^{1-\ep}} \prod_i \big( r_+^2+a_i^2 \big)\,.
\ee
With these, we are ready to discuss the two inequalities for higher dimensional rotating black holes.

\subsubsection{The lower bound}

The generalization of the lower bound in \eqref{thermo_inequality4D} for MP black holes turns out to be rather straightforward. We find
\bea
-\fft{D-4}{D-2} \fft{T}{S} &<& U \,, \qquad D=5,\, 7,\, 9,\, \cdots, \label{highKerroddlower}
\\
-\fft{D-3}{D-2} \fft{T}{S} &<& U \,, \qquad D=4,\, 6,\, 8,\, \cdots. \label{highKerrevenlower}
\eea
Note that for $D=4$ and $5$, the inequality sign ``$<$'' should be replaced by ``$\le$'', with the saturation occurring when all the angular momenta turned off. For $D\ge 6$, we find that we have to lower the bound determined by the static case further so that the bound is satisfied by all the angular momentum configuration, but with no particular saturation configuration. Compared to the parameter $1/2$ in $D=4$, the coefficients $(D-3)/(D-2)$ or $(D-4)/(D-2)$ are still purely numerical numbers of order unity.

\subsubsection{The upper bound}

The generalization of the upper bound in \eqref{thermo_inequality4D} for MP black holes turns out to be rather complicated. In particular, the upper bound with function $S$ alone cannot be established for $D\ge 6$. To see this, we start the discussion by setting all the angular momentum equal.  In this simplified case, we find that  \eqref{thermo_inequality4D} can be generalized to the following forms for $D\geq4$
\bea
U &\le& \frac{4^{\frac{1}{1-2 N}} (N-1) N^{\frac{1-N}{2 N-1}} \Sigma_{D-2}^{\frac{1}{2 N-1}} S^{\frac{2 N}{1-2 N}}}{\pi }\,, \quad D=5,\, 7,\, 9,\, \cdots, \label{highKerrodd}\\
U &\le& \frac{2^{-\frac{1}{N}-\frac{3}{2}} (2 N-1) \Sigma_{D-2}^{\frac{1}{2 N}} S^{-\frac{1}{2 N}-1}}{\pi  \sqrt{N}}\,, \quad D=4,\, 6,\, 8,\, \cdots. \label{highKerreven}
\eea
It recovers the previously-established $D=4$ case, which has only one angular momentum. For higher dimensions, there can be multiple angular momenta, depending on the dimensions. For example, in $D=5$, there are two angular momentum parameters $(a_1,a_2)$. We now show that the upper bound \eqref{highKerrodd} hold for general two angular momenta in $D=5$.  To be specific, we define
\bea
Y_1 &=& U+\fft13 \fft{T}{S}
= \frac{4 \left(a_1^4 a_2^4+2 a_1^4 a_2^2 r_+^2+2 a_1^2 a_2^4 r_+^2+5 a_1^2 a_2^2 r_+^4+a_1^2 r_+^6+a_2^2 r_+^6\right)}{3 \pi ^3 \left(a_1^2+r_+^2\right)^3 \left(a_2^2+r_+^2\right)^3}\,,
\\
Y_2 &=& U-\fft{\Sigma_3^{1/3}}{2\pi  S^{4/3}}
= \fft{Y_3}{3 \pi ^3 \left(a_1^2+r_+^2\right)^3 \left(a_2^2+r_+^2\right)^3}\,,
\eea
where
\bea
Y_3 &=& 5 a_1^4 a_2^4+3 r_+^6 \big(a_1^2+a_2^2\big)+20 a_1^2 a_2^2 r_+^4+9 a_1^2 a_2^2 r_+^2 \big(a_1^2+a_2^2\big) \nn \\
&&-3\times 2^{2/3} r_+^{4/3} \big(a_1^2+r_+^2\big)^{5/3} \big(a_2^2+r_+^2\big)^{5/3}-r_+^8\,.
\eea
It is easy to see that $Y_1\ge 0$ and the saturation appears when all angular momenta vanish, indicating that the $D=5$ Schwarzschild black hole saturates the lower bound. On the other hand, the sign of $Y_3$ is more difficult to determine. For simplicity, we set $r_+=1$ without loss of generality since we can measure $a_1$ and $a_2$ in unit of $r_+$. The temperature should be nonnegative, which provides a constraint $a_1 a_2 \le 1$. Here, we assume, without loss of generality, that $a_1$ and $a_2$ are both nonnegative. It is advantageous to introduce $\eta$ such that
\be
a_1 a_2 = 1-\eta\,,\qquad 0\le \eta\le 1\,.
\ee
Then we have $Y_3=Y_3(a_1,\eta)$. It is then easy to establish that for any fixed $\eta$, $Y_3$ has a maximum with $a_1=\sqrt{1-\eta}$, for which
\bea
Y_{3,\rm max} &=& 5 \eta ^4-38 \eta ^3+104 \eta ^2-3\times 2^{2/3}\, (2-\eta )^{10/3}-120 \eta +48\cr 
&=& -\xi ^2 \left(3\times 2^{2/3}\, \xi ^{4/3}-5 \xi ^2+2 \xi +4\right)\,.
\eea
where $\xi=(2-\eta)\in [1,2]$. It is straightforward to establish that the bracket quantity above is a monotonically decreasing function in the range $\xi \in [1,2]$, with the mininum value 0 when $\xi=2$. This therefore establishes that $Y_3$ is nonnegative and it approaches zero for the extremal black hole with two equal angular momenta.

For $D\ge 6$, we find that the quantity $U$ cannot be bounded by the quantity $S^\alpha$ with appropriate $\alpha$ and any numerical coefficient. We illustrate this in detail for $D=6$, where there are at most two angular momenta. The black hole can only be extremal when $a_1 a_2\ne 0$. Zero temperature requires
\be
a_2 = r_+ \sqrt{\fft{a_1^2 + 3 r_+^2}{a_1^2 - r_+^2}}\,,
\ee
which implies that $a_1$ can be arbitrarily large. We define
\be
Y_2 = U - \eta \fft{3\Sigma_{4}^{1/4}}{4\sqrt2 S^{5/4}}\,,
\ee
where $\eta=1$ corresponds to the bound established when all angular momenta are equal. We find that for large $a_1$, we have
\be
Y_2=\frac{3}{16 \pi ^3 a_1^2 r_+^3}-\frac{9 \eta }{16 \left(\sqrt[4]{2} \pi ^3 r_+^{5/2}\right)a_1^{5/2}}+O\left(\left(\frac{1}{a_1}\right)^4\right)\,.
\ee
This illustrate that no matter how large the coefficient $\eta$ is, $Y_2$ will turn to positive for sufficiently large $a_1$. It should be emphasized that this does not mean that $U$ is unbounded above; it simply cannot be bounded by the quantity $c/S^{5/4}$ for any constant value of $c$. We find that this is a general feature in $D\ge 6$ dimensions.

\section{Conclusion}
\label{sec:conclusion}

In this paper, we considered a general class of black holes with mass $M$ and conserved charges $Q_i$, which include angular momenta and electromagnetic charges. We proposed a sequence of inequalities \eqref{thermo_inequality4D} where the quantity $U$ is related to the inverse of the specific heat capacity $C_{Q_i}$ of fixed charges. The vanishing of $U$ is the thermodynamic phase transition point in that the positive $U$ is locally stable whilst the negative $U$ is locally unstable. Regardless of its positive or negative values, the sequence of inequalities \eqref{thermo_inequality4D} provides an entropy bound by the specific heat capacity, given by \eqref{bound}.

For special static black holes with $g_{tt}g_{rr}=-1$, we can prove the inequalities with suitable energy conditions. In particular, the lower bound requires SEC, whilst the DEC is sufficient to ensure the upper bound. We can also prove the upper bound for general static, spherically symmetric black holes in the extremal case, provided the DEC holds. For the more general black holes including rotations, the proof is unlikely at the current stage and we confirm the inequalities by examining a variety examples. We also studied the higher-dimensional generalizations and we expect that the bounds hold for static black holes. However, we explicitly verified with the MP black holes that there will not be an upper bound for $U$ by the entropy alone for $D\ge 6$ when rotation is involved.

About the entropy bounds, the most famous one is perhaps the Penrose inequality \cite{Bray:2003ns,Mars:2009cj}, which for stationary black holes, reads
\be
S\le 4\pi M^2\,.
\ee
Since then, a variety of inequalities involving black hole thermodynamic variables have been proposed, e.g.~\cite{Visser:1992qh,Dain:2011pi,Yang:2020ark,Yang:2022yye,Khodabakhshi:2022jot}.
These inequalities are all algebraic relations of the thermodynamic variables. What stands out in our new proposal is that it involves the derivative of the thermodynamic variables, namely the specific heat capacity of fixed charges. At first sight, we would naturally expect that the proof would be much more difficult since it is typically less clear how to relate a thermodynamic derivative term to the metric functions. Yet for the special static black holes, we presented a full proof of the bounds \eqref{bound}. The confirmation by this special class of black holes adds credibility of our general conjecture.

Having obtained the black hole bounds by the specific heat, it is natural to ask whether the bounds apply to other thermodynamic systems. Note that black hole thermodynamics is promoted from the black hole dynamics by the semiclassical quantum effects. To include all the involved physical fundamental constants which we have set to unity, the bounds of \eqref{bound} become
\be
\begin{cases}
S \le -\fft12 C_{Q_i}\,, \qquad\qquad\qquad\quad\,\,\,\,\,  C_{Q_i}<0\,,  \\
S \le \Big(\alpha \sqrt{\fft{c^5 \hbar}{k_B G}}\, T^{-1} C_{Q_i}\Big)^{\fft23}\,, \qquad  C_{Q_i}>0\,.
\label{bound2}
\end{cases}
\ee
For the locally stable systems, the bound above has explicit dependence of the fundamental constants $(\hbar,G,c)$, suggesting that this is the bound for quantum gravity. For the unstable systems, on the other hand, the coefficient is completely numerical, suggesting that the bound may apply to more general thermodynamic systems. It is of great interest to investigate this further.

\section*{Acknowledgement}
	
We are grateful to Run-Qiu Yang for useful discussions. This work is supported in part by the National Natural Science Foundation of China (NSFC) grants No.~12375052 and No.~11935009, and also by the Tianjin University Self-Innovation Fund Extreme Basic Research Project Grant No.~2025XJ21-0007.

\subsection*{Appendices}

\appendix

\section{The form of $f(r)$}
\label{app:ansatz f}

For the metric \eqref{metric1} with $\chi=0$ and the energy-momentum tensor $T^{\mu}_{\nu} = \text{diag}(-\rho, p_r, p_T, p_T)$, the $(t,t)$ component of the Einstein Equation gives
\be
\fft{1}{r^2} \fft{\dd}{\dd r} \Big[r \big(1-f(r) \big)\Big] = 8\pi \rho(r)\,.
\label{ode}
\ee
To solve this, we can integrate directly. A more physically intuitive approach is to define the ``Hawking-Geroch mass" $m(r)$ \cite{Hawking:1968qt,Geroch} via the relation:
\be
m(r) \equiv \fft{r}{2}\big(1-f\big)\,,
\label{mass}
\ee
which defines $m(r)$ as the function that makes the metric locally resemble a Schwarzschild metric. Substituting this definition into \eqref{ode} leads to
\be
\fft{\dd m}{\dd r} = 4\pi r^2 \rho(r)\,.
\label{dm-dr}
\ee
For energy density $\rho$ that falls off sufficiently fast, the solution is given by
\be
m(r) = M - \int_r^\infty 4\pi \tr^2 \rho(\tr) \dd\tr\,,\qquad f(r) = 1 - \fft{2M}{r} + \fft{8\pi}{r}\int_r^\infty \tr^2 \rho(\tr) \dd\tr\,,
\ee
where $M$ is the ADM mass and the last term of $f(r)$ can be denoted as $g(r, Q_i)$, since $\rho(r)$ only depends on the matter distribution, which can be a function of charges $Q_i$, but not the total mass of the spacetime.

\section{EMD black holes with general $N$}
\label{app:gen-N}

With the thermodynamic properties in \eqref{TS-singly}, we find that
\bea
Y_1 &=& \fft{N q}{8 \pi^2 \mu^3 (2 \mu +(4-N) q)} \Big(\fft{\mu +q}{\mu }\Big)^{-N} , \\
Y_2 &=& -\fft{\mu^{\fft{3N}{4} -3} (\mu +q)^{-N}}{4\pi^2 \big(2\mu +(4-N)q\big)}  \Big[ \mu ^{\frac{N}{4}+1} -(N-2)q \mu^\fft{N}{4} +2 \mu (\mu +q)^\fft{N}{4} \nn \\
&&+(4-N) q (\mu +q)^\fft{N}{4}\Big]\,.
\eea
Since $0<N<4$, we can find that \eqref{lowbound} holds automatically. For $N \le 2$, \eqref{upbound} also holds. We can set $\mu=\beta q$ to remove the dimension and simply focus on the dimensionless parameters $N$ and $\beta$. $\beta$ can take any positive value. The sign of $Y_2$ depends on the sign of the bracket term in $Y_2$, which can be rewritten as
\be
Z_1= \Big[-(N-2) \beta ^\fft{N}{4} +\beta ^{\fft{N}{4}+1} +2\beta(\beta +1)^\fft{N}{4} +(4-N) (\beta +1)^\fft{N}{4} \Big] q^{\fft{N}{4}+1},
\ee
and we hope to prove that $Z_1 \ge 0$. Note that we have the following relation
\be
Z_1 \ge Z_2= \Big[ \beta^{\fft{N}4} \big( 3(2+\beta)-2N \big)\Big] q^{\frac{N}{4}+1}.
\ee
It is easy to see that $Z_2>0$ for $N<3$. For $3<N<4$, $Z_2>0$ also holds for $\beta>\ft23$. Therefore, the undetermined parameter region reduces to $3<N<4$ and $0<\beta<\ft23$. However, $\beta$ and $N$ couple in $Z_1$ and it is hard to prove $Z_1>0$ analytically. The sign of $Z_1$ depends on the sign of the bracket term, which we call it $Z_3$. By plotting the picture of $Z_3$, we find that $Z_3\ge 0$ in the whole region. The saturation point is at $N=4,\,\beta=0$, which is precisely the extremal RN black hole. This completes our discussion of charged black holes in EMD.

\begin{figure}
	\centering
	\includegraphics[width=0.6\linewidth]{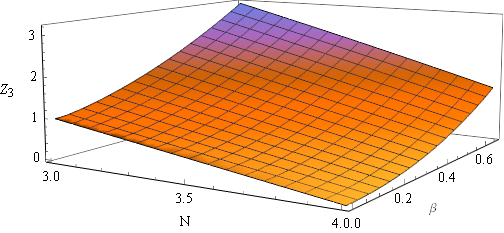}
	\caption{The 3D numerical picture of $Z_3>0$ in the parameter regions $3<N<4$ and $0<\beta<\ft23$.}
	\label{fig:genN}
\end{figure}

\section{The sign of $Y_3$ of the Kerr-dS black hole}
\label{app:KerrdS}

The variables of $Y_3$ are $r_+\,, g$ and $a$, but the value ranges of them are less clear. However, we must have positive temperature \eqref{TS-KerrdS}. Therefore, we can introduce a parameter $t\geq0$, which is defined as
\be
-a^2 g^2 r_+^2-a^2-3 g^2 r_+^4+r_+^2= r_+^2\big(1-3g^2r_+^2\big)t\,.
\ee
Then, we can replace the unit angular momentum $a$ by $t$, which is given by the following relation
\be
a=\frac{r_+ \sqrt{\big(1-3 g^2 r_+^2\big)\big(1-t\big)}}{\sqrt{1+g^2 r_+^2}}\,.
\label{KAds-a}
\ee
In order to get real $a$, we must have $0\le t\le 1$ and $1-3 g^2 r_+^2\geq 0$. For simplicity, we can set $\sqrt{1-t}=x$ in \eqref{KAds-a}, and $0\le x\le1$. Furthermore, the second constraint can be used to replace $r_+$, which allows us to avoid the trouble that meets the cosmic horizon. Therefore, we introduce the second parameter $y$, which is defined as $1-3 g^2 r_+^2=1-y^2$ and $0<y<1$. Using this new parameter, $r_+$ can be replaced by
\be
r_+=\fft{y}{\sqrt{3} g}\,.
\label{KAds-r0}
\ee
With \eqref{KAds-a} and \eqref{KAds-r0}, $Y_3$ can be rewritten as
\bea
Y_3 &=& -\fft{y^5}{9 \sqrt{3} g^5 (y^2+3)^{7/2}} \bigg[ 2 (y^2+3) \Big(3 x^2 (1-y^2)+y^2+3\Big)^2 \sqrt{-2 x^2 y^4+2 (x^2+1) y^2+6} \nn \\
&&+\sqrt{3 x^2 (1-y^2)+y^2+3} \Big(3 x^6 y^2 (y^2-3) (y^2-1)^3-3 x^4 (y^2-1)^2 (2 y^6-7 y^4-30 y^2+27) \nn \\
&&+x^2 (y^2+3)^2 (3 y^6-26 y^4+41 y^2-18)+(y^2+1) (y^2+3)^3\Big)  \bigg]\,.
\eea
The sign of $Y_3$ depends on the large bracket term in the above expression, which we call it $Y_4$. The analytical proof of the sign of $Y_4$ is also rather complicated. Therefore, we simply draw the numerical picture of $Y_4$ in the region $0\le x \le 1,\, 0<y<1$, and we find that $Y_4\ge 0$ in the whole region. At the point $x=1,\, y=0$, $Y_4=0$, which corresponds to $a=0,\, r_+=0$, and it is the pure de Sitter space. However, this still does not lead to the saturation of the upper bound since the exponent of the denominator is larger than that of the numerator.
\begin{figure}[tbh]
	\centering
	\includegraphics[width=0.6\linewidth]{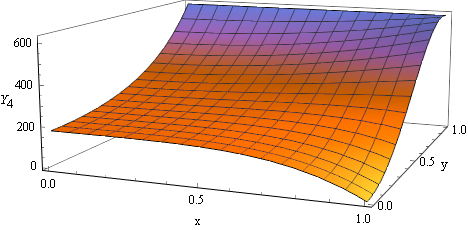}
	\caption{The numerical picture of $Y_4>0$ of Kerr-dS black hole in the parameter regions $0\le x\le 1$ and $0<y<1$.}
	\label{fig:KerrdS}
\end{figure}

\end{document}